# two-channel Compartmentalized Microfluidic Chip for Real-Time Monitoring of the Metastatic Cascade

Hilaria Mollica[1,2], Roberto Palomba[1], Rosita Primavera[1], Paolo Decuzzi[1,♣]

1. Laboratory of Nanotechnology for Precision Medicine, Italian Institute of technology, Via Morego 30, Genoa, 16163, Italy

2. DIBRIS, University of Genova, Via Opera Pia 13, Genoa, 16145, Italy

♣ Corresponding author: Dr. Paolo Decuzzi, paolo.decuzzi@iit.it

**KEYWORDS:** organ-on-a-chip, cell migration, real-time cell microscopy, cancer inflammation

**ABSTRACT**

Metastases are the primary cause of death in cancer patients. Small animal models are helping in dissecting some of key features in the metastatic cascade. Yet, tools for systematically analyze the contribution of blood flow, vascular permeability, inflammation, tissue architecture, and biochemical stimuli are missing. In this work, a microfluidic chip is designed and tested to replicate in vitro key steps in the metastatic cascade. It comprises two channels, resting on the same plane, connected via an array of rounded pillars to form a permeable micro-membrane. One channel acts as a vascular compartment and is coated by a fully confluent monolayer of endothelial cells, whereas the other channel is filled with a mixture of matrigel and breast cancer cells (MDA-MB-231) and reproduces the malignant tissue. The vascular permeability can be finely modulated by inducing pro-inflammatory conditions in the tissue compartment, which transiently opens up the tight junctions of endothelial cells. Permeability ranges from 1 µm/sec (tight endothelium) to 5 µm/sec (TNF-α at 50 ng/mL overnight) and up to ∼ 10 µm/sec (no endothelium). Fresh medium flowing continuously in the vascular compartment is sufficient to induce cancer cell intravasation at rates of 8 cells/day with an average velocity of ∼ 0.5 µm/min. On the other hand, the vascular adhesion and extravasation of circulating cancer cells require TNF-α stimulation. Extravasation occurs at lower rates with 4 cells/day and an average velocity of ∼ 0.1 µm/min. Finally, the same chip is completely filled with matrigel and the migration of cancer cells from one channel to the other is monitored over a region of about 400 µm. Invasion rates of 12 cells/day are documented upon TNF-α stimulation. This work demonstrates that the proposed compartmentalized microfluidic chip can efficiently replicate in vitro, under controlled biophysical and biochemical conditions, the multiple key steps in the cancer metastatic cascade.

## INTRODUCTION

It is well accepted that metastases and disease recurrence are the main causes of death in cancer patients.[1] The ability of malignant cells to enter the blood stream, abandoning the primary tumor mass; disseminate through the vascular network, searching for a new homing tissue; adhere to the vascular walls, resisting hemodynamic forces; extravasate at a distant site, opening the endothelial barrier; and, eventually, migrate away from the blood vessels, infiltrating a new tissue; is crucial in the formation of metastatic niches.[2-4] This stepwise sequence of events is regulated by a multitude of biophysical and biochemical processes, including alterations of cell polarity, cytoskeletal and nuclear architecture, and expression of membrane receptors. For instance, the intravasation of tumor cells is supported by the well-known epithelial-to-mesenchymal transition (EMT), which involves the disruption of intercellular adhesion, cell polarity and the overexpression of specific cell-matrix adhesion molecules.[5, 6] Inside the vascular network, tumor cells, which are then called circulating tumor cells (CTCs), can interact and cluster with blood cells, such as platelets, to elude immune surveillance and enhance longevity in the blood stream. Also, CTCs can interact with the vascular walls establishing transient and firm adhesion events with the endothelial cells, mimicking what leukocytes do at sites of inflammation[4]. The small fraction of tumor cells surviving the vascular environment can progressively infiltrate the new homing tissue and modify the local microenvironment to create suitable conditions for engraftment and proliferation. Although the mechanisms are yet to be fully elucidated, establishing metastatic niches is not a random process but is affected by the local microenvironment, vascular architecture as well as by the type and location of the original malignant mass.[7-11]

Different microscopy techniques, including conventional confocal fluorescent microscopy and intravital video microscopy, have been employed to monitor the fate of individual cancer cells and the progressive formation of metastatic niches. In this context, non-mammalian model organisms, such as zebrafishes and drosophilas, and small rodents, such as mice and rats, have been used to recapitulate the metastatic evolution of different tumor types.[12, 13] Non-mammalian model organisms allow conventional confocal microscopy to follow in vivo the dynamics of individual CTCs. Also, through genetic modification, these models can be efficiently used to test specific biological hypotheses. For instance, drosophilas and zebrafishes have been used to elegantly reveal the pivotal role of matrix metalloproteinase secretion and tumor microenvironment in supporting cancer cell migration and spreading to distant organs.[14-16] Although more demanding in terms of equipment and animal handling, intravital microscopy techniques have been developed by multiple laboratories to observe the vascular dynamics of CTCs and tissue infiltration in real time. For instance, Winkler and collaborators focused on the formation of brain metastases by observing the evolution of individual CTCs over a time window ranging from minutes to months.[17, 18] The group of Robert Hoffman has developed preclinical models of cancer metastasis, optical probes and microscopy techniques for assessing single cancer cell dynamics in different organs, including the lungs[19], pancreas[20] and prostate.[21] Protocols have been also developed to image the formation and evolution of clusters of CTCs.[22] Although animal models offer a more authentic representation of the key biophysical and biochemical features regulating cancer metastases, it is difficult to control, accurately and independently, the many governing parameters. For this reason, systematic analyses on the temporal and spatial evolution of CTCs can be more efficiently performed solely using in vitro assays.

Cell migration across biological barriers has been traditionally studied using a Boyden chamber assay, which was originally employed for mimicking leukocyte chemotaxis and, then, applied to study tumor cell invasion.[23] In this system, a porous membrane separates two different compartments, which contain the cells and media of interest. Despite its simplicity, the Boyden chamber assay does not allow one to monitor in real time cell migration from one chamber to the other, given the difference in focal planes; and, more importantly, cannot support fluid flow, which affects cell dynamics, cytoskeletal architecture and receptor expression. In order to address these limitations, microfluidic-based assays have been developed over the last few years with diverse applications. For instance, cell infiltration was studied by Chaw and colleagues, who fabricated a multi-step microfluidic device where cells are forced to squeeze within tiny and long orifices filled with extracellular matrices.[24] A similar approach was also employed by others to study cancer cell migration under chemical gradients and electric fields.[25] The group of Roger Kamm realized a microfluidic platform comprising a central and two lateral channels, separated by an intermediate chamber, to study cancer cell migration, invasion and extravasation.[26-29] Other groups mostly focused on CTC adhesion to endothelial cells under flow, upon stimulation by specific chemokines and pro-inflammatory molecules.[30, 31] Yet, an experimental set-up for reproducing the multiple, key events in the metastatic cascade is missing.

In this work, a compartmentalized microfluidic device is proposed, which comprises two micro-channels running in parallel and connected by a micro-membrane realized in the lateral walls (**Figure.1**). Differently from a Boyden chamber, the device allows one to monitor simultaneously the dynamics of cells within the two different compartments and across the micro-membrane. This is microfabricated to include a series of pillars forming openings smaller than 3 μm, which separate the two channels into two different compartments. Endothelial cells are deposited in the vascular compartment forming a confluent layer over the micro-membrane, whereas an extracellular matrix enriched with different cell types is deposited within the tissue compartment. The device can be used for studying cell intravasation, vascular transport and adhesion, extravasation and invasion, thus helping to address key steps in the complex metastatic cascade.

**MATERIAL AND METHODS**

**Fabrication of the double microfluidic chip.** The fabrication of the microfluidic chip device involved two lithographic steps, as detailed in the sequel. Following protocols previously detailed by the authors[32], the chip was obtained using a top-down fabrication approach.

Two optical masks, one negative for the two parallel channels and one positive for the micropillars membrane, were realized using a laser writing machine (DLW6000). The AZ5214E (Microchem) photoresist was employed due to its capability to work both in negative and positive processes. It was spin-coated at 4,000 rpm on a Cr surface, then the resulting sample was baked a 110 ˚C for 60 s to clean the resist. The mask including the micromembrane was impressed using a mask-aligner on the resist at 80 mJ. The solvent AZ726MF (Microchem) was used as a developer for the resist. The impressed pattern was transferred from the resist to the Cr mask by using a commercial Cr etchant (Chrome etch 18, OSC OrganoSpezialChemie GmbH). The residual resist layer was then removed from the unexposed surface via acetone cleansing and sonication. An ICP-RIE Bosh process (Si 500, Sentech Instruments GmbH) was performed to transfer the micropillar membrane over the Si, etching down by 5 µm. In the Bosh process, each cycle consists of a deposited phase and an etching phase. The obtained Si wafer with the impressed micromembrane was used then for transferring the channels. The fabrication method was the same used for the pillars except for the resist, which was here used in reversal mode (i.e. negative mode). AZ5214E was spin-coated on the residual Cr layer and baked at 110˚C for 60 s. After a realignment phase between the channels mask on the micromembrane, the photoresist was exposed at 95 mJ. Since the resist was used in reversal mode, the wafer required an additional baking phase at 120˚C for 120 s and an exposure phase at 300 mJ. Then, developing was performed as for the micromembrane. At this point, both pillars and channels are impressed on the Cr layer, whereas the pillars only are on the Si wafer. A second ICP-RIE Bosh etching was performed down to 50 µm, which is the actual height of the channels. Then, both pillars and channels were transferred on the Si layer. Before replicating this geometry on a polydimethylsiloxane (PDMS) layer, an anti-stiction film of 1H,1H,2H,2H-Perfluoro-octyl-trichloro-silane (125µl, Alfa Aesar) was deposited on the silicon template by using a desiccator for 1h. The PDMS replica was obtained by using a mixture of a base and curing agent with a ratio (w/w) 1:10 (Sylgard 182, Dow Corning). This solution was casted on the silicon template and baked in an oven at 60 °C, overnight for 15h. The PDMS replica underwent an oxygen ($O_2$) plasma treatment (Pressure = 0.5 mBar, Power = 20 W, Time = 20 s; Plasma System Tucano, Gambetti) and bonded to a glass coversheet (No. 1.5H, Deckaläser). A biopsy punch (OD = 1 mm, Miltex) was used to create inlet and outlet ports. The resulting microfluidic chip had a rectangular cross section of w = 210 µm (width), h = 50 µm (height), and l = 2.7 cm (port-to-port length), as shown in **Figure 1a**. The micropillar membrane had a length l = 500 µm and a width w = 25 µm. Pillars are separated by a 3 µm gap.

**Electron microscopy imaging.** Microfluidic chips were analyzed via Scanning Electron Microscopy (JSM-6490LV, JEOL and Helios Nanolab 650, FEI Company™). SEM images were acquired after cutting the chip, bonded on the glass slide, with a blade. Then cells were fixed with 0.2% of Glutaraldehyde in cacodylate 0.1 M buffer solution and dried with an ethanol solution. The PDMS was pretreat with 15 nm of gold and then imaged by secondary electrons imaging (SEI) mode. Low magnification and high magnification SEM images were obtained with accelerating voltage of 15 and 5 kV, respectively.

**Permeability experiments in the microfluidic chip.** Microfluidic chips were autoclaved at 120 °C for sterilization. Then they were dried to remove the water from channels and placed in an incubator overnight (37 °C, humidity > 95%). Matrigel 8-12 mg/mL (Sigma Aldrich) was maintained in ice, then it was mixed with Eagle's minimum essential medium (EMEM) (ATCC©, USA) to yield a final matrix concentration of 4-6 mg/mL. 50 ng/mL of TNF-α were introduced in the matrigel, while on ice to prevent gelation. This allowed the homogeneous dispersion of TNF-α in a still liquid matrigel solution. Then, the liquid mix of matrigel with 50 ng/mL TNF-α was introduced in the chamber. Gelation occurred in 5 minutes at 37 °C. TNF-α was gradually released from the extravascular compartment. Next, the vascular channel was filled with 20 µg/mL of a fibronectin solution (Sigma Aldrich) and incubated for 1h at 37°C. Human Umbilical Vein Endothelial Cells (HUVEC) were cultured in endothelial growth medium according to the manufacturer's guidelines (cells and culture media were purchased from PromoCell, USA). Before seeding into the microfluidic chip, HUVECs were cultured, washed, detached, counted and concentrated at $6\times10^6$ cell/mL. Cells were used up to passage 6 (P6). For both the vascular and the extravascular compartments, micropipette tips on the inlet side were filled with 200 µl of culture media, whereas the tips on the outlet side were left empty. Media was changed every day. Chips were used 1 day after seeding. The vascular channel was connected to a syringe pump (Harvard Pump 11 Elite, Harvard Apparatus) by a polyethylene tubing (BTPE-50, ID = 0.58 mm, OD = 0.97 mm, Instech Laboratories). During the permeability tests, 40 kDa FITC-Dextran (Sigma Aldrich) was added up to a concentration of 0.5 µg/mL. For all the experiments, solutions were flowed at a physiological shear rate, $\sigma = 20$ s$^{-1}$, which corresponds to a volumetric flow rate of 100 nL/min. Dextran permeability was detected using an epi-fluorescent inverted microscope (Leica 6000, objective 10 x, 0.22 N.A.) and the ImageJ software, using two Region of Interest (ROI), one fixed in the vascular channel and the other in the extravascular channel. The formula, for calculating the permeability coefficients P, is reported in **Figure.2b**. Specifically, the equation $P = (I_f - I_i)w/[(I_i - I_b)\Delta t]$ was used, where P is the permeability (µm/s), $I_f$ is the total fluorescent intensity in the two ROIs at the final time, $I_i$ is the total fluorescent intensity in the two ROIs at the initial time. $I_b$ is the total fluorescent intensity in the extravascular ROI at the time zero. $\Delta t$ is the time interval between the two analyzed frames, and w is the width of the vascular channel.

**Cancer cell intravasation model.** Breast cancer MDA-MB-231 cells were used at $1.5\times10^7$ cell/mL, embedded in the matrigel solution, and then introduced in the extravascular channel. After matrigel gelation, the vascular channel was filled with 20 µg/mL of a fibronectin solution. Then, the chips were incubated for 1h at 37°C. Then, vascular endothelial cells HUVECs were inserted at $6\times10^6$ cells/mL. Cancer cells were imaged a 6, 21, 24 and 30 h using a fluorescent inverted microscope (Leica, objective 10x). Any cancer cell with its full body in the vascular compartment was considered an intravasated cell. Time-lapse movies were acquired for the whole duration of the experiment, while media was continuously introduced with a syringe pump at Q = 50 nL/min.

**Cancer cell transport and adhesion under flow model.** The vascular channel was filled with 20 µg/mL of a fibronectin solution and incubated for 1h at 37°C. Then, HUVECs were inserted at $6\times10^6$ cells/mL. The extravascular channel was filled with a matrigel matrix and 50 ng/mL of a TNF-α solution. After reaching cell confluency, the chip was placed on the stage of an epi-fluorescence inverted microscope performing the adhesion experiments. The working fluid was injected into the chip using a syringe pump (33 Dual, Harvard apparatus). After the trypsinization, the cancer cells were incubated for 30 min with CM-DiI, at 37°C (0.5%, Thermofisher) according to the manufacture's protocol. Then, cells were washed 3 times with PBS 1x (GIBCO) to remove the dye excess. Finally, cells were re-suspended in EMEM without FBS, at $1\times10^6$ cells/mL. After each rolling experiment, the chip was washed with PBS to remove non-adherent cancer cells. Tumor cells ($1\times10^6$ cells/mL) were introduced via a syringe pump on the HUVEC monolayer. The inlet port of the chip was connected to the syringe pump through a polyethylene tube (BTPE-50, ID = 0.58 mm, OD = 0.97 mm, Instech Laboratories). The interaction of tumor cells with HUVECs was recorded for 15 minutes for each experiment. Two flow rates Q were imposed via the syringe pump, namely, 50 and 100 nL/min. [33].

**Cancer cell extravasation model.** The vascular channel was filled with 20 µg/mL of a fibronectin solution and incubated for 1h at 37°C, HUVECs were inserted at $6\times10^6$ cells/mL. The extravascular channel was filled with a solution of matrigel, FITC-Dextran 4 kDa and TNF-α 100 µg/mL. After 24 h, MDA-MB-231 cells were infused ($1\times10^6$ cells/mL) on the endothelial cells by using a syringe pump at the flow rate of 50 nL/min, for 15 minutes. The extravasated cancer cells were imaged at a Leica microscope at 3, 6, 21, 24 and 30h. Cancer cells with a full body in the extravascular compartment were considered extravasated cells. Time-lapse movies were acquired for the whole duration of the experiment, while media was continuously introduced with a syringe pump at Q = 50 nL/min.

**Cancer cell invasion model.** Both channels in the device were filled with matrigel. One compartment was filled with MDA-MB-231 cancer cells ($1.5 \times 10^7$), while the other was filled with FITC-Dextran 4 kDa and TNF-α 50 ng/mL. Migrating cancer cells were imaged at a Leica microscope at 3, 6, 21, 24 and 30h. Cancer cells with a full body in the opposite compartment were considered migrating cells. Time-lapse movies were acquired for the whole duration of the experiment, while micropipettes tips were filled with 200 μl of cells growth media for both compartments.

**Cell tracking.** For the intravasation, extravasation and invasion experiments, images were acquired every 4 minutes for 30 h. During the whole experiment chips were kept under controlled environmental conditions (5% $CO_2$ and 37°C), using an Oko lab Cage Incubation System mounted on the time-lapse microscope. Movies were acquired with a 20 x objective. Cell dynamic was detected using time-lapse imaging set-up, including an on-stage incubator (Eclipse Ti-E, Nikon) and a sCMOS camera (Andor Zyla). The confocal fluorescent images in the two different channels were obtained using *Split- Channels* and recombining the images using *Merge Channels*.

**Immunofluorescence staining and image acquisition.** At the end of the experiments, both vascular and extravascular channels were carefully washed with PBS (Invitrogen) and fixed with 4% paraformaldehyde (PFA) (ChemCruz, Santa Cruz, Biotechnologies, USA) for 15 minutes at room temperature. After washing twice with PBS, channels were filled with a 0.3% Triton X in PBS solution for 10 min at 4°C to allow cellular membrane permeation. Next, cells were incubated with 20% goat serum in PBS solution for 30 min at 4°C and then human endothelial Cadherins were targeted for 3 hours using anti-human VE-cadherin antibody (Ms anti-human VE-cadherin, 1:200, Abcam), while ICAM-1 adhesion molecules were targeted for 3 hours using anti-human ICAM-1 antibody (Rb anti-human ICAM-1, 1:200, Abcam). Afterwards, the chips were incubated with a green fluorescent labeled secondary antibody (anti-mouse 488, 1:500, Abcam) and with a red fluorescent labeled secondary antibody (anti-rabbit 467, 1:500, Abcam), for 50 min at 4 °C. Cell nuclei were stained with DAPI (5 mg/mL, Invitrogen) while F-Actin cytoskeleton filaments were stained in green using phalloidin, according to the supplier instructions (Alexa Fluor® phalloidin, life technologies). Images were acquired using a confocal microscope (Nikon A1).

## RESULTS

**The compartmentalized microfluidic device.** A custom designed polydimethylsiloxane (PDMS) microfluidic device was used to reproduce the key steps in the metastatic process. The device consists of two parallel channels divided by a permeable membrane of rounded pillars (**Figure.1**). The channels present a total length of 2.7 cm (from inlet to outlet), a height of 50 µm and a width of 210 µm. The equivalent hydraulic diameter of the channels is comparable in size to large capillaries, arterioles and venules. The permeable membrane is placed in the center of the channels and is 500 µm long. The separation distance between the two channels, across the pillar membrane, is equal to~ 3 µm, which is sufficient to compartmentalize the two channels while still allowing molecules and cells to diffuse through. Therefore, the two compartments can be independently filled at the occurrence with different matrices (collagen, matrigel, hyaluronic acid, and combinations thereof) and cells (endothelial, cancer, stromal, immune cells and so on). In **Figure.1a,** a schematic of the microfluidic devices is presented together with electron microscopy images revealing the details of the micro-membrane. Specifically, the first electron micrograph shows the central portion of the device with the arrays of pillars constituting the micro-membrane.

The second electron micrograph returns the shape and separation distance between adjacent pillars within the micro-membrane. These are slender structures, with an aspect ratio larger than 2.5, presenting a rounded shape to enhance lateral mechanical stability. In **Figure.1b,** two representative confocal fluorescent images document the compartmentalization of the device in two different channels. On the left, endothelial cells (HUVECs) confluently cover the upper, lower and lateral walls of the vascular channel (cell nuclei in blue – DAPI; VE-cadherin in green – FITC). On the right, breast cancer MDA-MB-231 cells populate a matrigel matrix deposited in the extravascular channel (cell membrane stained in red – CM-DiI). Notice that the endothelial cells adhere over the entire exposed surface in the vascular compartment allowing the formation of a pervious vessel with a tubular shape, whose permeability can be modulated at the micro-membrane. Additional details on the membrane and channels' cross sections are provided by the electron micrographs of **Figure.1c**. In the right image, cells can be observed distributed within the channels, whereas the left image offers a detailed view of the micropillars in the membrane.

**Vascular permeability at the micro-membrane.** In order to test the functionality of the endothelial barrier, the vascular channel was seeded with endothelial cells while the other channel was filled with matrigel to mimic the extracellular matrix. Endothelial cells spontaneously cover the walls, including the permeable micro-membrane. The diffusion of a green fluorescent tracer (FITC–Dextran 40 kDa) from the vascular to the extravascular compartment was analyzed to retrieve values for the vascular permeability P, under different operating conditions. Specifically, endothelial cells were stimulated with the pro-inflammatory molecule TNF-α to modulate intercellular adhesion and, thus, vascular permeability. The fluorescent images in **Figure.2a** document the permeation of the tracer in the extravascular space at 3 different time points, namely 5, 15 and 30 min post infusion. Dextran molecules were observed to flow into the vascular channel, permeate across the micro-membrane, and diffuse into the matrigel matrix, as also depicted in the **Supporting Movies**.**1**, **2** and **3**. The diffusion of the fluorescent molecule is inversely proportional to the tightness of the endothelial junctions, which are loosened by the treatment with TNF-α. In the absence of endothelial cells (no HUVECs), the tracer easily flows into the extravascular compartment returning permeability values as high as 9.25 ± 3.96 µm/s. On the contrary, the presence of a continuous endothelial layer dramatically reduces the vascular permeability of the 40 kDa tracer to 1.01 ± 0.34 µm/s. Furthermore, the stimulation with TNF-α, which was added in the extravascular compartment at 50 ng/mL, affects the tight junctions (**Supporting Figure.1**) of the endothelial cells and increases the permeability of the tracer up to 5.43 ± 2.42 µm/s. Note that the TNF-α concentration is low enough to avoid any toxic and irreversible effect on the cells but sufficient to alter adhesive molecules expression (**Supporting Figure.2**). In all cases, the obtained values are slightly higher with respect to in vivo data, but are in line with in vitro data reported by other authors.[34] The chart in **Figure.2b** gives the permeability values as well as the formula employed to extract these values from the experimental data. Indeed, these results demonstrate the ability to modulate the vascular permeability in the device by properly stimulating confluent endothelial cells. Permeability experiments were also conducted at lower flow rates, namely 50 nL/min (**Supporting Figure.3**). Importantly, only a minor difference in permeability was observed for the two different flow rates.

**Modeling the intravasation of cancer cells.** After assessing the vascular permeability of the device under different flows conditions and TNF-α stimulations (**Supporting Figure.4**), the transport of breast cancer cells (MDA-MB-231) across the endothelialized micro-membrane into the vascular compartment was considered. Tumor cells were mixed with matrigel and infused into the extravascular compartment, whereas endothelial cells were seeded and cultured in the vascular channel. During the whole experiment, cell culture medium was continuously infused on the vascular side with a flow rate of Q = 50 nL/min, which is typical for microvascular flow.[35] Despite the high density of matrigel, the fresh medium and nutrients in the vascular channel were sufficient to attract cancer cells. Intravasation events were observed over time and were only accounted for when the whole cell body was found in the vascular compartment. **Figure.3a** presents a schematic of the model and fluorescent microscopy images at the two compartments taken at 0, 6, 21, 24 and 30 h post cell seeding. These images document a progressive infiltration of the breast cancer cells (cell membrane in red – CM-DiI) into the vascular compartment, which is confluently covered by endothelial cells (cell nuclei in blue – DAPI). It should be noted that in some of the microfluidic chips a small number of HUVECs migrated into the extravascular compartment. Indeed, HUVECs feel the presence of cancer cells and their cytokines, and respond to their biological stimuli by migrating into extravascular channel. Moreover, matrigel does contain 5.0-7.5 ng/mL of vascular endothelial growth factor (VEGF) and traces of matrix metalloproteinases (MMP) that could contribute to this process. The number of MDA-MB-231 was calculated by observing the tumor cells for 30h at the time-lapse microscope, as in the **Supporting Movie.4.** This shows tumor cells migrating into the matrigel matrix, crossing the endothelial barrier, and reaching the vascular compartment. In **Supporting Figure.5** the trajectory and velocity of an individual intravasated tumor cell are reported. The analysis was performed with a time-lapse microscope returning an average speed of about 0.008 µm/sec, and a total variation in speed from 0.0005 to 0.018 µm/sec. In **Supporting Figure.6** details on the morphology of intravasating cancer cells are provided. As compared to HUVECs, the migrating cancer cells appear elongated. In **Figure.3b**, the number of intravasated tumor cells is plotted at specific time points, returning the averaged values 2.70 ± 2.31 at 6h; 5.77 ± 3.30 at 21h; 8.28 ± 4.30 at 24h; 9.55 ± 4.24 at 30h. In **Figure.3c** and in **Supporting Figure.6,** confocal fluorescent images of the device depict the intravasated cancer cells (red) across the micro-membrane. MDA-MB-231 cells are stained in red with CM-DiI, the nuclei of the HUVECs are stained in blue with DAPI and the F-Actin filaments of both cells are stained in green with Alex Fluor 488 phalloidin (for **Supporting Figure.6** notice that nuclei of both HUVECs and MDA-MB-231 are stained in blue with DAPI). It should be here emphasized that similar images and data on cell migration velocity could have not been taken in a Boyden chamber assay.

**Modeling vascular transport and adhesion of cancer cells.** After intravasation into the vascular compartment, tumor cells are transported by the blood flow until vascular adhesion occurs at secondary sites. To model this process, MDA-MB-231 cells were infused in the endothelialized vascular channel at 50 and 100 nL/min. To modulate the adhesion propensity of the cancer cells, HUVECs were either treated with 50 ng/mL of TNF-α (inflamed endothelium) or left in their basal state (no inflammation). The stimulation with TNF-α enhances the expression of specific molecules, such as the vascular cell adhesion, proteins and integrins that mediate cell adhesion. As in the previous section, TNF-α (50 ng/mL) was added in the extravascular compartment, mimicking an inflammatory stimulus originating deep in the tissue and eventually reaching the endothelial barrier. **Figure.4a** reports representative fluorescent images showing breast cancer cells (red – CM-DiI) adhering onto HUVECs (blue – DAPI), under the two tested flow rates. The full movies documenting the vascular transport and adhesion of the breast cancer cells are provided in **Supporting Movies.5** and **6**. The number of adhering cancer cells was quantified and normalized by the total number of injected cells ($n_{inj}=10^6$) and the area of the region of interest (ROI) (**Figure.4b**). At low flow rates, the number of adhering cancer cells was equal to 12.94 ± 4.47 for unstimulated conditions (-TNF-α) and 29.75 ± 4.19 for stimulated conditions (+TNF-α). At higher flow rates, the number of adhering cancer cells decreased to 8.38 ± 3.14 for unstimulated conditions (-TNF-α) and 21.00 ± 4.38 for stimulated conditions (+TNF-α). As expected, CTCs would adhere on the inflamed endothelium almost 3-times more than on the healthy vasculature. Differently, CTCs would more stably roll on the healthy endothelium rather than on the inflamed vasculature.

**Modeling the extravasation of cancer cells.** After establishing stable adhesion with the blood vessel walls, tumor cells can migrate towards the extravascular space infiltrating the healthy tissue.[36] MDA-MB-231 cell extravasation from the vascular compartment was evaluated tracking cells for 30h on a time-lapse microscope. In this configuration, tumor cells are required to cross the endothelial barrier. As such, HUVECs (blue – DAPI) were seeded in the vascular compartment to form a confluent cell layer, while MDA-MB-231 (red – CM-DiI) were infused in the same compartment at 50 nL/min. The extravascular channel was filled with a mix of matrigel, TNF-α and FITC-Dextran (4 kDa). The Dextran 4 kDa tracer was included to monitor the transport of small molecules, such as TNF-α, from the tissue to the vascular compartment. The transmigration of MDA-MB-231 into the matrigel matrix was then observed over time. Representative fluorescent microscopy images of the extravasated cancer cells are shown in **Figure.5a** at different time points. **Figure.5b** reports the absolute number of extravasated cells, namely, 1.42 ± 0.78 at 3h, 2.42 ± 0.97 at 6h, 3.71 ± 0.75 at 21h, 4.42 ± 1.39 at 24h, 5.00 ± 1.91 at 30h. In **Supporting Figure.7,** the trajectory and the velocity of an individual extravasated tumor cell are reported. The analysis was performed with a time-lapse microscope. The average speed of the tumor cell is 0.002 µm/sec, with an overall variation between 0.013 and 0 µm/sec. **Figure.5c** and **Supporting Figure.8** report confocal fluorescent images of the device depicting the extravasated cancer cells across the micro-membrane. MDA-MB-231 cells are stained in red with CM-DiI, the nuclei of the HUVECs are stained in blue with Dapi and the F-Actin filaments of both cells are stained in green with Alexa Fluor 488 phalloidin (for **Supporting Figure.8** notice that nuclei of both HUVECs and MDA-MB-231 are stained in blue with DAPI). **Supporting Movie.7** shows tumor cells migrating into the matrigel matrix, crossing the micro-membrane, and reaching the extravascular compartment.

**Modeling the tissue invasion of cancer cells.** After extravasation, cancer cells colonize the secondary sites by penetrating deeper into the tissue and forming metastatic niches. In this step, cancer cells interact with local microenvironment, including stromal cells, that would facilitate invasion by releasing chemokines, MMPs and other molecules[37].

To model this step of the metastasis cascade, the device was filled with a mix of matrigel, TNF-α and FITC-Dextran 4 kDa in one channel, and MDA-MB-231 cells embedded in matrigel in the other channel. Cancer cell migration from one side to the other was evaluated by tracking cells for up to 30h, using a time-lapse microscope. **Figure.6a** shows the schematic of the chip and representative fluorescent images taken using an inverted microscope. The absolute number of migrating cells was charted in **Figure.6b** returning the values at 0.50 ± 0.81 at 3h, 2.42 ± 1.30 at 6 h, 10.42 ± 3.78 at 21h, 12.71 ± 4.03 at 24h and 14.71 ± 3.13 at 30 h. In **Supporting Figure.9**, the trajectory and velocity of an individual tumor cell are reported. The average cell speed is 0.005 μm/sec, with a total variation ranging between 0.011 and 0 μm/sec. Cancer cells move along the TNF-α gradient crossing the channels from one side to the other through the permeable micro-membrane. Indeed, for this study, no endothelial cells were included. A drop of cell culture medium was added to the inlet and to the outlet ports to prevent matrigel drying. **Figure.6c** and **Supporting Figure.10** report confocal fluorescent images depicting the migrating cancer cells across the micro-membrane. MDA-MB-231 cells are stained in red with CM-DiI, the nuclei are stained in blue with DAPI and the F-Actin filaments are stained in green with Alex Fluor 488 phalloidin.

## DISCUSSION AND CONCLUSIONS

A compartmentalized microfluidic chip has been realized comprising two channels acting as the vascular compartment, coated by a confluent layer of endothelial cells, and the tissue compartment, filled by a matrigel matrix enriched with breast cancer cells. A micro-membrane, made out of an array of rounded pillars, separates the two channels and realizes the chip compartmentalization. Multiple key steps in the cancer metastasis process – intravasation, vascular transport and adhesion, extravasation and tissue invasion – have been reproduced in this compartmentalized microfluidic chip under controlled conditions. Importantly, as the two channels rest on the same plane, the trans-membrane dynamics of molecules and individual cancer cells can be monitored in real-time over several hours, using time-lapse fluorescent microscopy. This allows one to accurately quantify the rates of cell intravasation, adhesion, extravasation and invasion as well as to extract relevant information on cell morphology and biophysics.

No specific biological and biophysical information on organs and vascular districts were included in this version of the work as the main focus is on demonstrating the ability to control in real-time key steps in the metastatic cascade. Nonetheless, the size of the vascular chamber can be readily modified during the fabrication process and the flow rate can be accurately modulated by programming a syringe pump. By tuning these two parameters, the flow conditions in different vascular districts can be readily reproduced. As such, four conditions can be, for instance, recapitulated in vitro, as those found in 5 – 10 µm capillaries, with a mean blood velocity of 0.01 cm/sec; 10 – 50 µm arterioles, with a mean blood velocity of 0.6 cm/sec; 10 – 70 µm venules, with a mean blood velocity of 0.8 cm/sec; and 2 mm collecting lymphatic vessels, with a mean blood velocity of 10 cm/sec. Furthermore organ specific conditions can be integrated by using different types of endothelial cells within the vascular chamber, and multi-cellular mixture in the extravascular chamber. Considering flow conditions of venules, arterioles or large capillaries, in this work, MDA-MB-231 cells were observed to migrate towards the vascular compartment with a rate of about 8 cells per day. This process is relatively rapid, with an average cell velocity of 0.5 µm/min. Notice, however, that the total number of cancer cells dispersed within the tissue compartment (matrigel matrix) and potentially available for intravasation is equal to about 4,000. In other words, the intravasation process is characterized by a low efficiency with only 0.2% of the tumor cells migrating from the malignant mass towards the vasculature each day. Indeed, this number can be affected by a variety of factors including the initial cell density. On the other hand, it was observed that cell extravasation requires stimulation with the pro-inflammatory molecule TNF-$\alpha$, which favors the opening of the endothelial barrier and acts as a chemoattractant on tumor cells. Still, the rate of cell deposition within the extravascular space is of only 4 cells per day. This step in the metastatic cascade is even less efficient than intravasation and average cell velocities of about 0.1 µm/min were measured. TNF-$\alpha$ was also used for studying the invasion of cancer cells deep into the tissue. In this case, a migration rate of 12 cells per day over a 400 µm long region is observed. Is it important to highlight that these values of cancer cell velocity are in agreement with other in vitro and in vivo data available in literature as documented in the **Supporting Table.1**. This quantitatively data on cell migration average speed, confirms that cancer metastasis is a highly inefficient process and specific conditions and stimuli are needed to support it. Intravasation, vascular adhesion, extravasation and invasion are all affected by a variety of biophysical and biochemical stimuli, including the local hemodynamic conditions, vascular permeability, expressions of adhesion molecules, availability of pro-inflammatory stimuli chemoattractant molecules, density and type of the extracellular matrix. The proposed two-channel compartmentalized microfluidic chip allows one to control accurately all the above parameters and provide a useful tool to systematically characterize the metastatic process dissect new biological mechanism or identify new anticancer therapies.

**SUPPORTING INFORMATIONS**

The Supporting Informations provide: quantification of VE-cadherin expression on HUVECs borders (Figure S1); visualization of ICAM-1 adhesion molecules (Figure S2); vascular permeability at micro-membrane at flow rate Q= 50 nL/min (Figure S3); vascular permeability at micro-membrane at flow rate Q= 100 nL/min ( Figure S4); single cell tracking and instantaneous velocity of intravasated cancer cell (Figure S5); a representative confocal pictures of intravasated cancer cells (Figure S6); single cell tracking and instantaneous velocity of extravasated cancer cell (Figure S7); a representative confocal pictures of extravasated cancer cells (Figure S8); single cell tracking and instantaneous velocity of migrating cancer cell (Figure S9); a representative confocal pictures of migrating cancer cells (Figure S10);   Velocity of different cancer cells during the metastatic steps (Table S1). Permeability studies in a compartmentalized microfluidic device (Movies S1, S2 and S3); cancer cell intravasation (Movie S4); Cancer cell adhesion under flow (Movies S5, S6); cancer cell extravasation (Movie S7) and cancer cell invasion (Movie S8).

**ACKNOWLEDGMENTS**

This project was partially supported by the European Research Council, under the European Union's Seventh Framework Programme (FP7/2007-2013)/ERC grant agreement no. 616695, by the Italian Association for Cancer Research (AIRC) under the individual investigator grant no. 17664, and by the European Union's Horizon 2020 research and innovation programme under the Marie Sk lodowska-Curie grant agreement no. 754490. Authors thank the Staff of the Clean Room and Nikon Center Facilities of the Italian Institute of Technology.

**REFERENCES**

1. Steeg, P.S., *Targeting metastasis.* Nat Rev Cancer, 2016. **16**(4): p. 201-18 DOI: 10.1038/nrc.2016.25.

2. Nguyen, D.X., P.D. Bos, and J. Massague, *Metastasis: from dissemination to organ-specific colonization.* Nat Rev Cancer, 2009. **9**(4): p. 274-84 DOI: 10.1038/nrc2622.

3. Gupta, G.P. and J. Massagué, *Cancer metastasis: building a framework.* Cell, 2006. **127**(4): p. 679-695 DOI: 10.1016/j.cell.2006.11.001.

4. Wirtz, D., K. Konstantopoulos, and P.C. Searson, *The physics of cancer: the role of physical interactions and mechanical forces in metastasis.* Nature Reviews Cancer, 2011. **11**(7): p. 512 DOI: 10.1038/nrc3080.

5. Gregory, P.A., A.G. Bert, E.L. Paterson, S.C. Barry, A. Tsykin, G. Farshid, M.A. Vadas, Y. Khew-Goodall, and G.J. Goodall, *The miR-200 family and miR-205 regulate epithelial to mesenchymal transition by targeting ZEB1 and SIP1.* Nature cell biology, 2008. **10**(5): p. 593 DOI: 10.1038/ncb1722. Epub 2008 Mar 30.

6. Macara, I.G. and L. McCaffrey, *Cell polarity in morphogenesis and metastasis.* Phil. Trans. R. Soc. B, 2013. **368**(1629): p. 20130012 DOI: 10.1098/rstb.2013.0012.

7. Paget, S., *The distribution of secondary growths in cancer of the breast.* The Lancet, 1889. **133**(3421): p. 571-573 DOI: https://doi.org/10.1016/S0140-6736(00)49915-0.

8. Ewing, J., *Neoplastic diseases; a treatise on tumors*, in *cdl; americana*. 1922, University of California Libraries: Philadelphia London, W. B. Saunders company p. 987-1031.

9. Cristofanilli, M., D.F. Hayes, G.T. Budd, M.J. Ellis, A. Stopeck, J.M. Reuben, G.V. Doyle, J. Matera, W.J. Allard, and M.C. Miller, *Circulating tumor cells: a novel prognostic factor for newly diagnosed metastatic breast cancer.* Journal of clinical oncology, 2005. **23**(7): p. 1420-1430 DOI: 10.1200/JCO.2005.08.140.

10. Azevedo, A.S., G. Follain, S. Patthabhiraman, S. Harlepp, and J.G. Goetz, *Metastasis of circulating tumor cells: favorable soil or suitable biomechanics, or both?* Cell adhesion & migration, 2015. **9**(5): p. 345-356 DOI: 10.1080/19336918.2015.1059563.

11. Yu, M., S. Stott, M. Toner, S. Maheswaran, and D.A. Haber, *Circulating tumor cells: approaches to isolation and characterization.* The Journal of cell biology, 2011. **192**(3): p. 373-382 DOI: doi: 10.1083/jcb.201010021.

12. Stuelten, C.H., C.A. Parent, and D.J. Montell, *Cell motility in cancer invasion and metastasis: insights from simple model organisms.* Nat Rev Cancer, 2018. **18**(5): p. 296-312 DOI: 10.1038/nrc.2018.15.

13. van Marion, D.M., U.M. Domanska, H. Timmer-Bosscha, and A.M. Walenkamp, *Studying cancer metastasis: existing models, challenges and future perspectives.* Critical reviews in oncology/hematology, 2016. **97**: p. 107-117 DOI: 10.1016/j.critrevonc.2015.08.009.

14. Miles, W.O., N.J. Dyson, and J.A. Walker, *Modeling tumor invasion and metastasis in Drosophila.* Dis Model Mech, 2011. **4**(6): p. 753-61 DOI: 10.1242/dmm.006908.

15. Beaucher, M., E. Hersperger, A. Page-McCaw, and A. Shearn, *Metastatic ability of Drosophila tumors depends on MMP activity.* Dev Biol, 2007. **303**(2): p. 625-34 DOI: 10.1016/j.ydbio.2006.12.001.

16. Cock-Rada, A.M., S. Medjkane, N. Janski, N. Yousfi, M. Perichon, M. Chaussepied, J. Chluba, G. Langsley, and J.B. Weitzman, *SMYD3 promotes cancer invasion by epigenetic upregulation of the metalloproteinase MMP-9.* Cancer Res, 2012. **72**(3): p. 810-20 DOI: 10.1158/0008-5472.CAN-11-1052.

17. Kienast, Y., L. von Baumgarten, M. Fuhrmann, W.E. Klinkert, R. Goldbrunner, J. Herms, and F. Winkler, *Real-time imaging reveals the single steps of brain metastasis formation.* Nat Med, 2010. **16**(1): p. 116-22 DOI: 10.1038/nm.2072.

18. Osswald, M., E. Jung, F. Sahm, G. Solecki, V. Venkataramani, J. Blaes, S. Weil, H. Horstmann, B. Wiestler, M. Syed, L. Huang, M. Ratliff, K. Karimian Jazi, F.T. Kurz, T.

Schmenger, D. Lemke, M. Gommel, M. Pauli, Y. Liao, P. Haring, S. Pusch, V. Herl, C. Steinhauser, D. Krunic, M. Jarahian, H. Miletic, A.S. Berghoff, O. Griesbeck, G. Kalamakis, O. Garaschuk, M. Preusser, S. Weiss, H. Liu, S. Heiland, M. Platten, P.E. Huber, T. Kuner, A. von Deimling, W. Wick, and F. Winkler, *Brain tumour cells interconnect to a functional and resistant network.* Nature, 2015. **528**(7580): p. 93-8 DOI: 10.1038/nature16071.

19. Kimura, H., K. Hayashi, K. Yamauchi, N. Yamamoto, H. Tsuchiya, K. Tomita, H. Kishimoto, M. Bouvet, and R.M. Hoffman, *Real-time imaging of single cancer-cell dynamics of lung metastasis.* J Cell Biochem, 2010. **109**(1): p. 58-64 DOI: 10.1002/jcb.22379.

20. Bouvet, M., J. Wang, S.R. Nardin, R. Nassirpour, M. Yang, E. Baranov, P. Jiang, A.R. Moossa, and R.M. Hoffman, *Real-time optical imaging of primary tumor growth and multiple metastatic events in a pancreatic cancer orthotopic model.* Cancer Res, 2002. **62**(5): p. 1534-40.

21. Zhang, Y., M. Toneri, H. Ma, Z. Yang, M. Bouvet, Y. Goto, N. Seki, and R.M. Hoffman, *Real-Time GFP Intravital Imaging of the Differences in Cellular and Angiogenic Behavior of Subcutaneous and Orthotopic Nude-Mouse Models of Human PC-3 Prostate Cancer.* J Cell Biochem, 2016. **117**(11): p. 2546-51 DOI: 10.1002/jcb.25547.

22. Garona, J. and D.F. Alonso, *Urokinase Exerts Antimetastatic Effects by Dissociating Clusters of Circulating Tumor Cells-Letter.* Cancer Res, 2016. **76**(16): p. 4908 DOI: 10.1158/0008-5472.CAN-16-0119.

23. Albini, A. and R. Benelli, *The chemoinvasion assay: a method to assess tumor and endothelial cell invasion and its modulation.* Nat Protoc, 2007. **2**(3): p. 504-11 DOI: 10.1038/nprot.2006.466.

24. Chaw, K.C., M. Manimaran, E.H. Tay, and S. Swaminathan, *Multi-step microfluidic device for studying cancer metastasis.* Lab Chip, 2007. **7**(8): p. 1041-7 DOI: 10.1039/b707399m.

25. Li, J., L. Zhu, M. Zhang, and F. Lin, *Microfluidic device for studying cell migration in single or co-existing chemical gradients and electric fields.* Biomicrofluidics, 2012. **6**(2): p. 24121-2412113 DOI: 10.1063/1.4718721.

26. Chung, S., R. Sudo, P.J. Mack, C.R. Wan, V. Vickerman, and R.D. Kamm, *Cell migration into scaffolds under co-culture conditions in a microfluidic platform.* Lab Chip, 2009. **9**(2): p. 269-75 DOI: 10.1039/b807585a.

27. Chen, M.B., J.M. Lamar, R. Li, R.O. Hynes, and R.D. Kamm, *Elucidation of the Roles of Tumor Integrin beta1 in the Extravasation Stage of the Metastasis Cascade.* Cancer Res, 2016. **76**(9): p. 2513-24 DOI: 10.1158/0008-5472.CAN-15-1325.

28. Chen, M.B., J.A. Whisler, J.S. Jeon, and R.D. Kamm, *Mechanisms of tumor cell extravasation in an in vitro microvascular network platform.* Integr Biol (Camb), 2013. **5**(10): p. 1262-71 DOI: 10.1039/c3ib40149a.

29. Jeon, J.S., S. Bersini, M. Gilardi, G. Dubini, J.L. Charest, M. Moretti, and R.D. Kamm, *Human 3D vascularized organotypic microfluidic assays to study breast cancer cell extravasation.* Proc Natl Acad Sci U S A, 2015. **112**(1): p. 214-9 DOI: 10.1073/pnas.1417115112.

30. Thompson, T.J. and B. Han, *Analysis of adhesion kinetics of cancer cells on inflamed endothelium using a microfluidic platform.* Biomicrofluidics, 2018. **12**(4): p. 042215 DOI: 10.1063/1.5025891.

31. Song, J.W., S.P. Cavnar, A.C. Walker, K.E. Luker, M. Gupta, Y.C. Tung, G.D. Luker, and S. Takayama, *Microfluidic endothelium for studying the intravascular adhesion of metastatic breast cancer cells.* PLoS One, 2009. **4**(6): p. e5756 DOI: 10.1371/journal.pone.0005756.

32. Manneschi, C., R. Pereira, G. Marinaro, A. Bosca, M. Francardi, and P. Decuzzi, *A microfluidic platform with permeable walls for the analysis of vascular and extravascular*

*mass transport.* Microfluidics and Nanofluidics, 2016. **20**(8): p. 113 DOI: 10.1007/s10404-016-1775-5.

33. Mollica, H., A. Coclite, M.E. Miali, R.C. Pereira, L. Paleari, C. Manneschi, A. DeCensi, and P. Decuzzi, *Deciphering the relative contribution of vascular inflammation and blood rheology in metastatic spreading.* Biomicrofluidics, 2018. **12**(4): p. 042205 DOI: 10.1063/1.5022879. eCollection 2018 Jul.

34. Albelda, S.M., P.M. Sampson, F.R. Haselton, J. McNiff, S. Mueller, S. Williams, A. Fishman, and E. Levine, *Permeability characteristics of cultured endothelial cell monolayers.* Journal of Applied Physiology, 1988. **64**(1): p. 308-322 DOI: 10.1152/jappl.1988.64.1.308.

35. Popel, A.S. and P.C. Johnson, *Microcirculation and hemorheology.* Annu. Rev. Fluid Mech., 2005. **37**: p. 43-69 DOI: 10.1146/annurev.fluid.37.042604.133933.

36. van Zijl, F., G. Krupitza, and W. Mikulits, *Initial steps of metastasis: cell invasion and endothelial transmigration.* Mutation Research/Reviews in Mutation Research, 2011. **728**(1): p. 23-34 DOI: 10.1016/j.mrrev.2011.05.002

37. Martin, T.A., L. Ye, A.J. Sanders, J. Lane, and W.G. Jiang, *Cancer invasion and metastasis: molecular and cellular perspective*, in *Madame Curie Bioscience Database [Internet]*. 2013, Landes Bioscience.

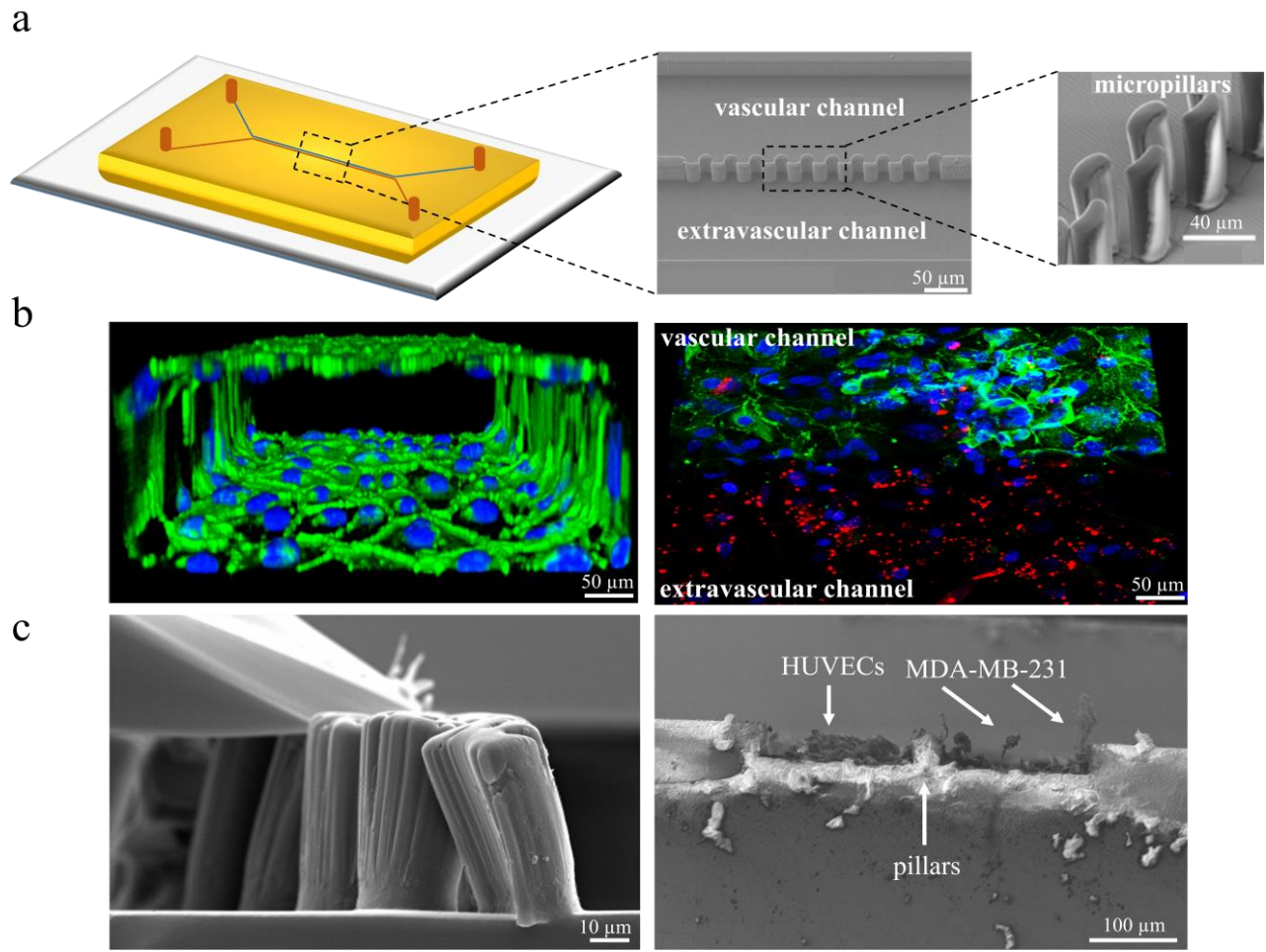

**Figure 1. The compartmentalized microfluidic device. a.** Schematic of the double channel chip (length = 2.7 cm, width = 210 µm; height = 50 µm) with the vascular channel in blue and the extravascular channel in red (left); scanning electron microscopy micrographs showing the extravascular and vascular channels; and the micro-membrane made out of micropillars, (right). **b.** Representative confocal fluorescent images of a confluent HUVEC monolayer in the vascular compartment (left) and MDA-MB-231 breast cancer cells mixed in a matrigel layer in the extravascular compartment (right). Cell nuclei are stained in blue with DAPI, VE-Cadherin proteins are stained in green, cellular membrane is stained in red with CM-DIL. **c.** Scanning electron microscopy micrographs showing details of the micro-membrane and its pillars (left), and cross section of the vascular channel filled with HUVECS and extravascular compartment filled with MDA-MB-231 cells (right).

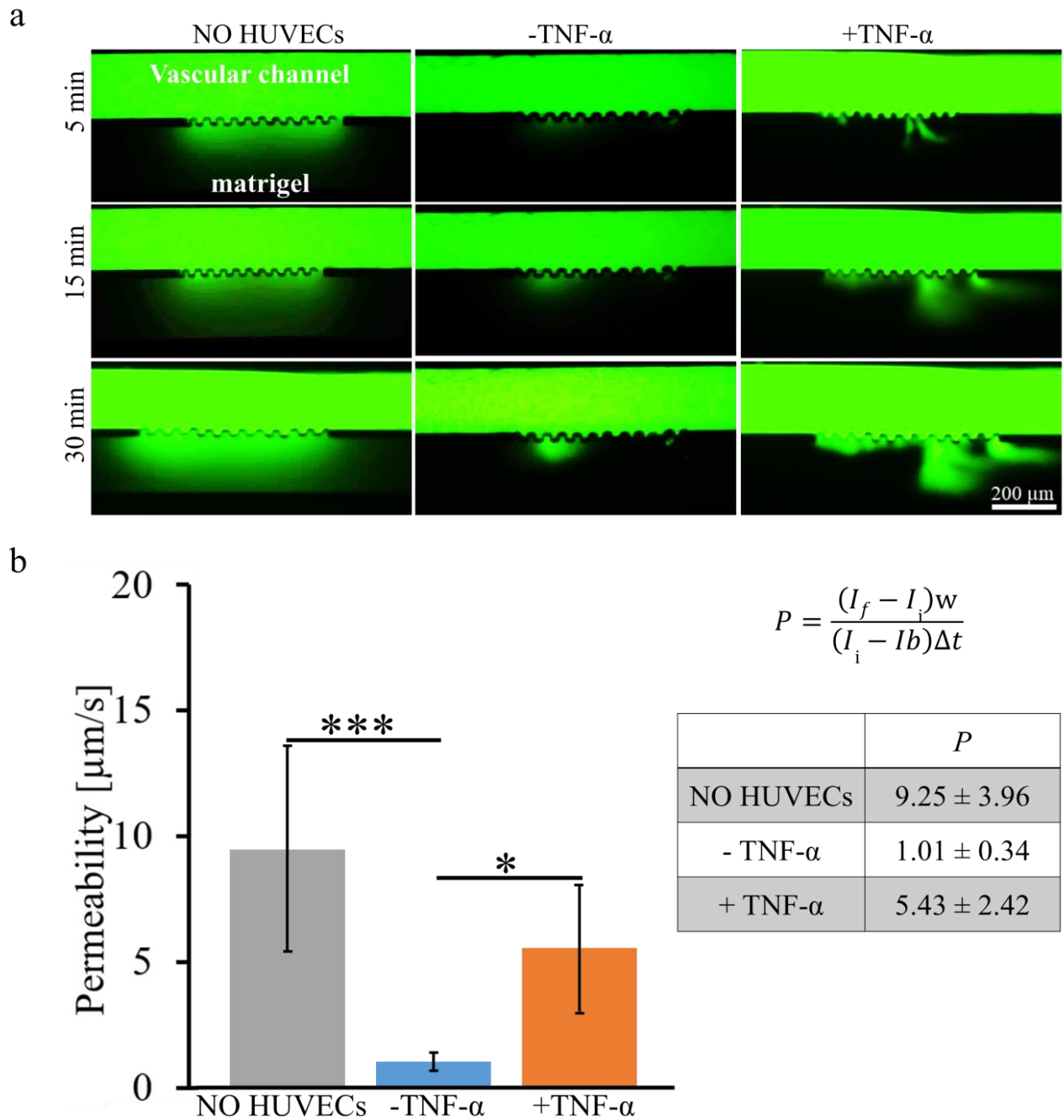

**Figure 2. Vascular permeability at the micro membrane. a.** Representative fluorescent images of free FITC-Dextran (40 kDa) diffusing (Q = 100 nL/min) in the vascular channel with an healthy (- TNF-α); inflamed endothelium (+ TNF-α) and with no endothelial cells (NO HUVECs) (TNF-α treatments were performed at 50 ng/mL for 12h). **b.** Vascular permeability coefficients. Formula for calculating the permeability (right). Data are plotted as mean + SD, n = 5. Statistical analysis ANOVA. * denotes statistically significant difference p<0.05. *** denotes statistically significant difference p<0.01.

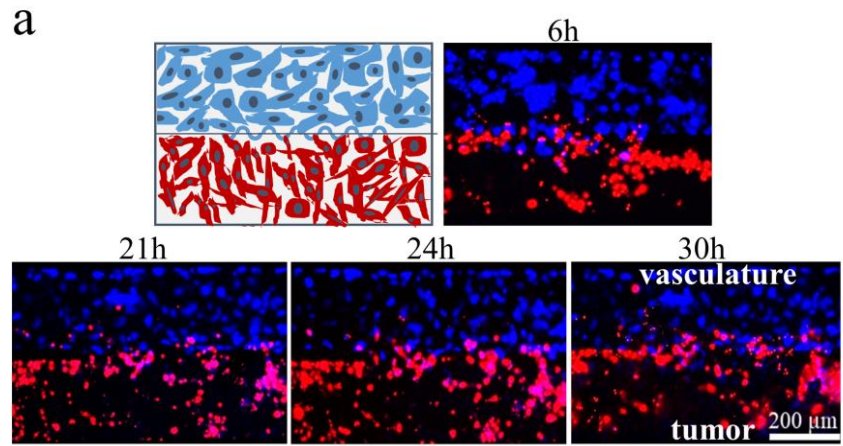
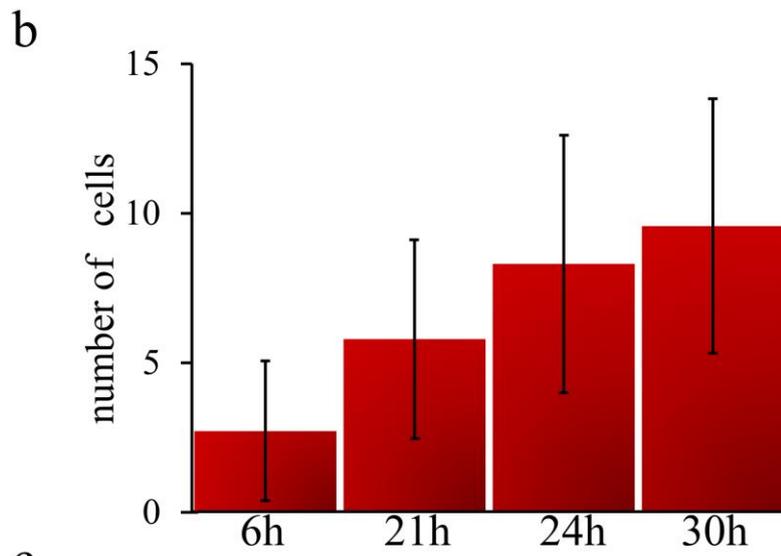
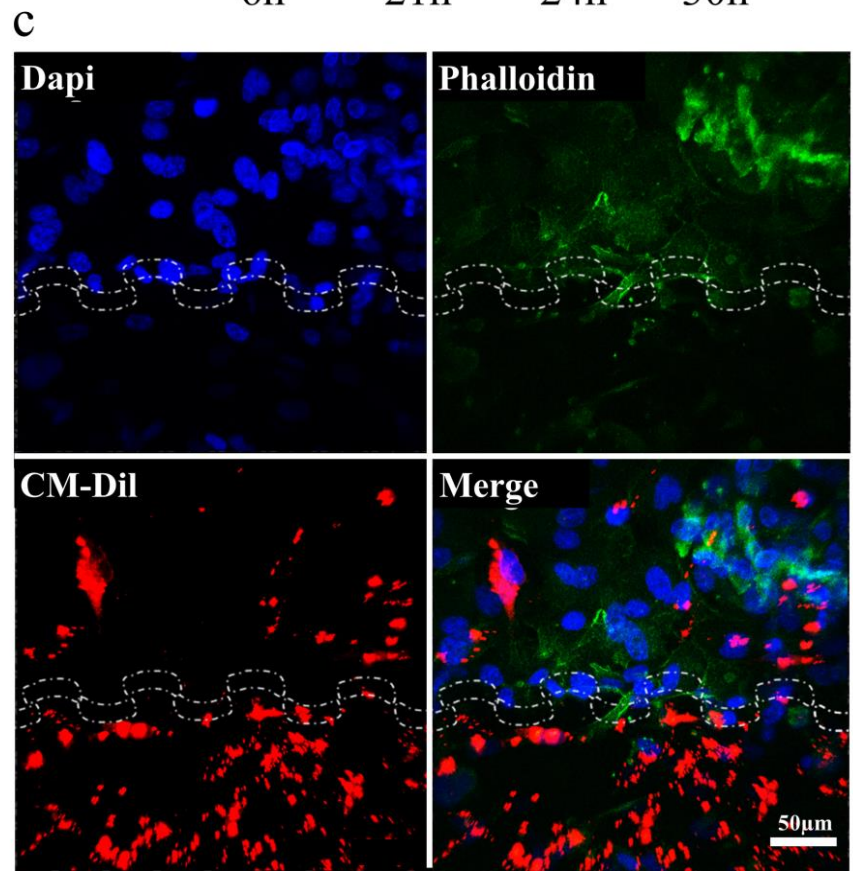

**Figure 3. Modeling the intravasation of cancer cells. a.** A schematic (top left) of the model and fluorescent microscopy images of the two compartments, taken at 6, 21, 24 and 30h. The images depict cancer cells (cell membrane labeled in red with CM-Dil) intravasated from the extravascular compartment filled with a matrigel matrix to the vascular compartment covered by a confluent layer of HUVECs (cell nuclei stained in blue with DAPI). **b.** Quantification of the number of intravasated cancer cells up to 30h. Data are plotted as mean + SD, n=8. **c.** Confocal fluorescent images into different channels were obtained using *Split-Channels* and recombining the images using *Merge-Channels*. HUVECs nuclei are stained in blue with DAPI, F-Actin is stained in green with phalloidin.

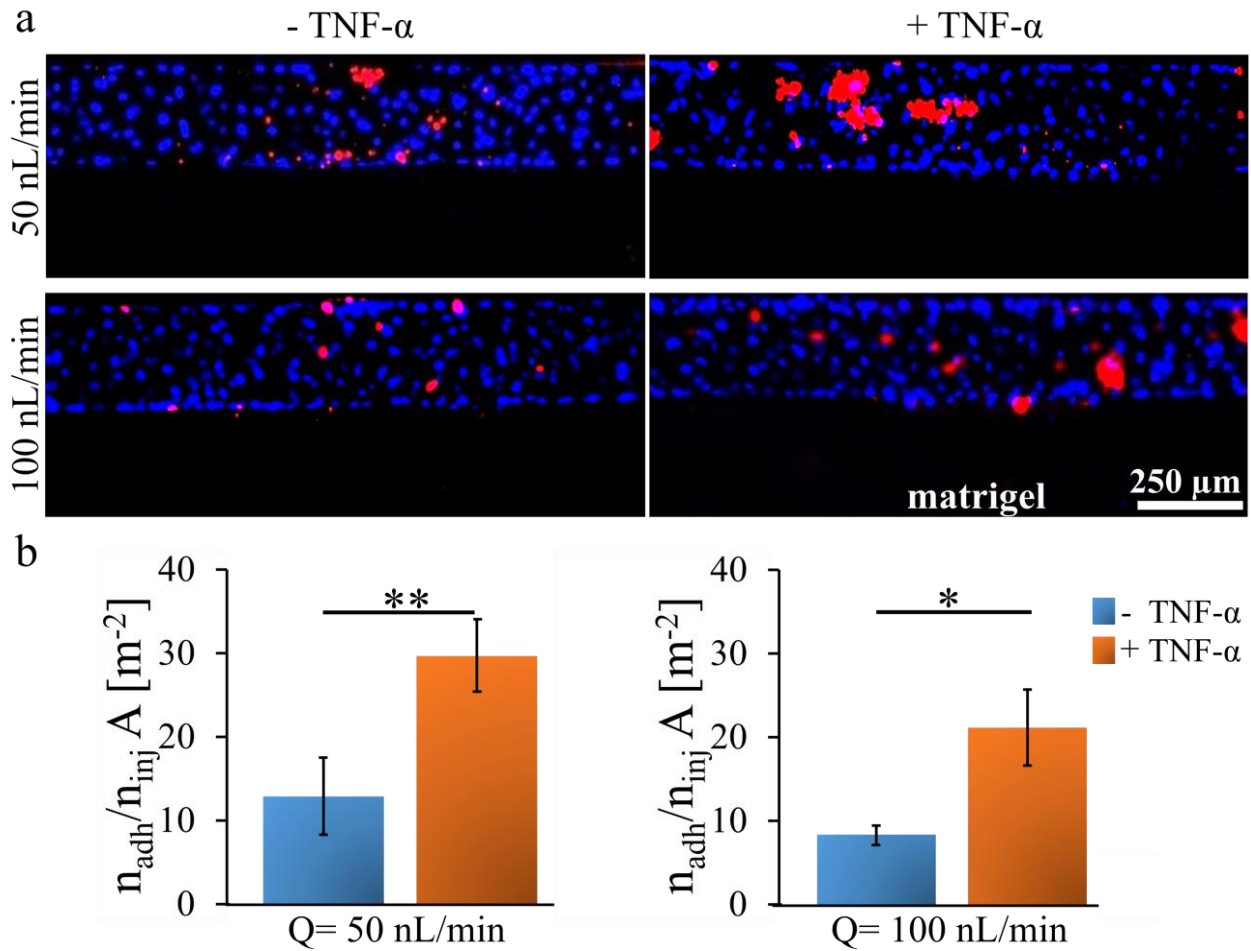

**Figure 4. Modeling the vascular transport and adhesion of cancer cells. a.** Representative fluorescent microscopy images depicting breast cancer cells (cell membrane labeled in red with CM-DIL) adhering over a confluent HUVEC monolayer (cell nuclei stained in blue with DAPI). HUVECs are either inflamed for 12h with TNF-α (50 ng/mL) (+ TNF-α) or not inflamed (- TNF-α). **b.** Normalized number of adhering cancer cells for flow rate Q = 50 nL/min (bottom left) and Q = 100 nL/min (bottom right). Data are plotted as mean + SD, n=3. Statistical analysis ANOVA. * denotes statistically significant difference $p<0.05$. ** denotes statistically significant difference $p<0.01$.

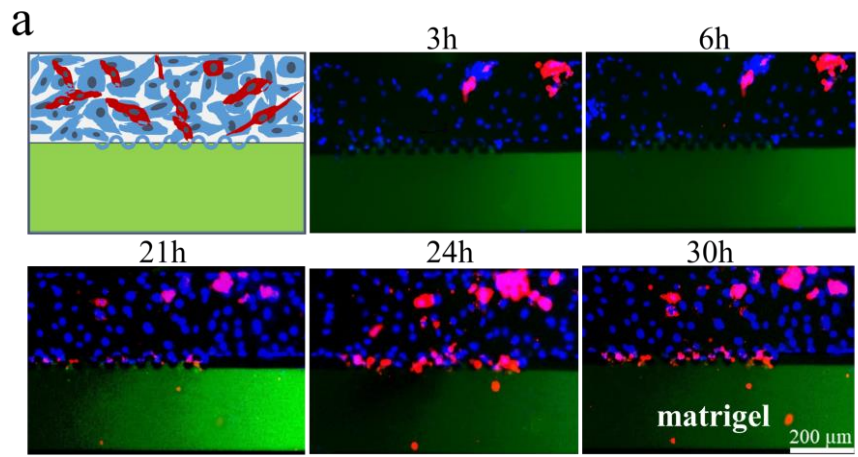

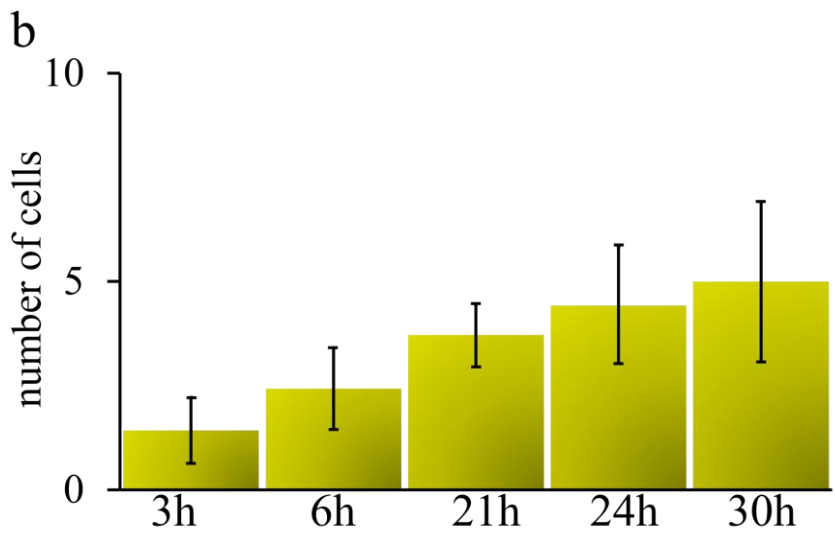

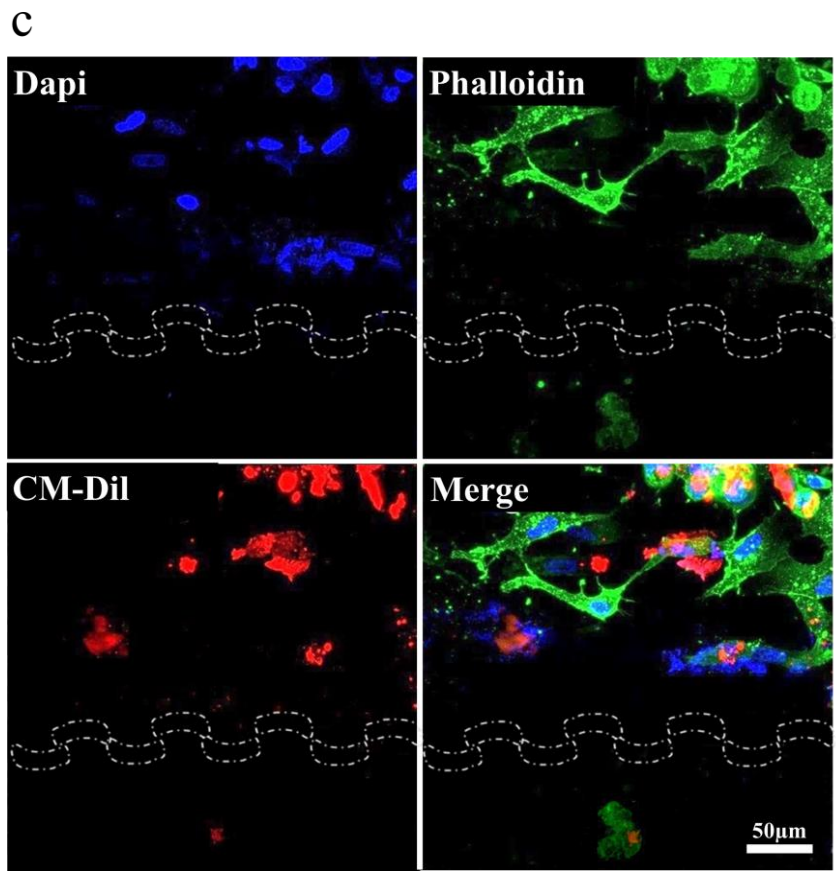

**Figure 5. Modeling the extravasation of cancer cells.** A schematic (top left) of the model and fluorescent microscopy images of the two compartments taken at 3, 6, 21, 24 and 30h. Images depict cancer cells (cell membrane labeled in red with CM-Dil) infiltrating from the vascular compartment, filled with HUVECs (cell nuclei stained in blue with DAPI), to the extravascular compartment filled with matrigel. **b.** Quantification of the number of extravasated cancer cells up to 30h. Data are plotted as mean + SD, n=8. **c.** Confocal fluorescent images into different channels were obtained using *Split- Channels* and recombining the images using *Merge-Channels*. HUVECs nuclei are stained in blue with DAPI, F-Actin is stained in green with phalloidin.

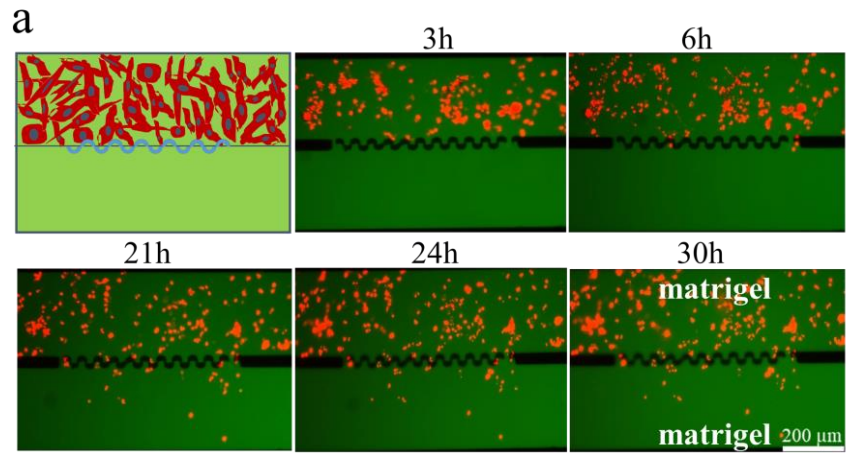
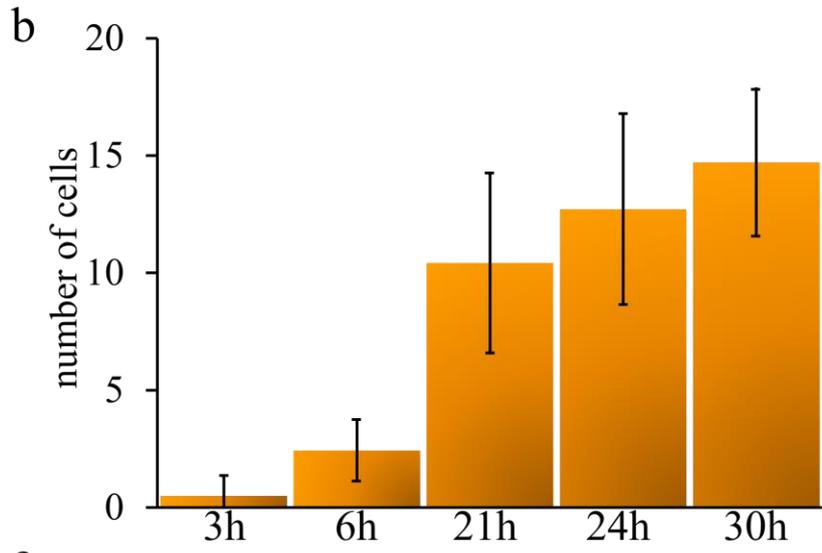
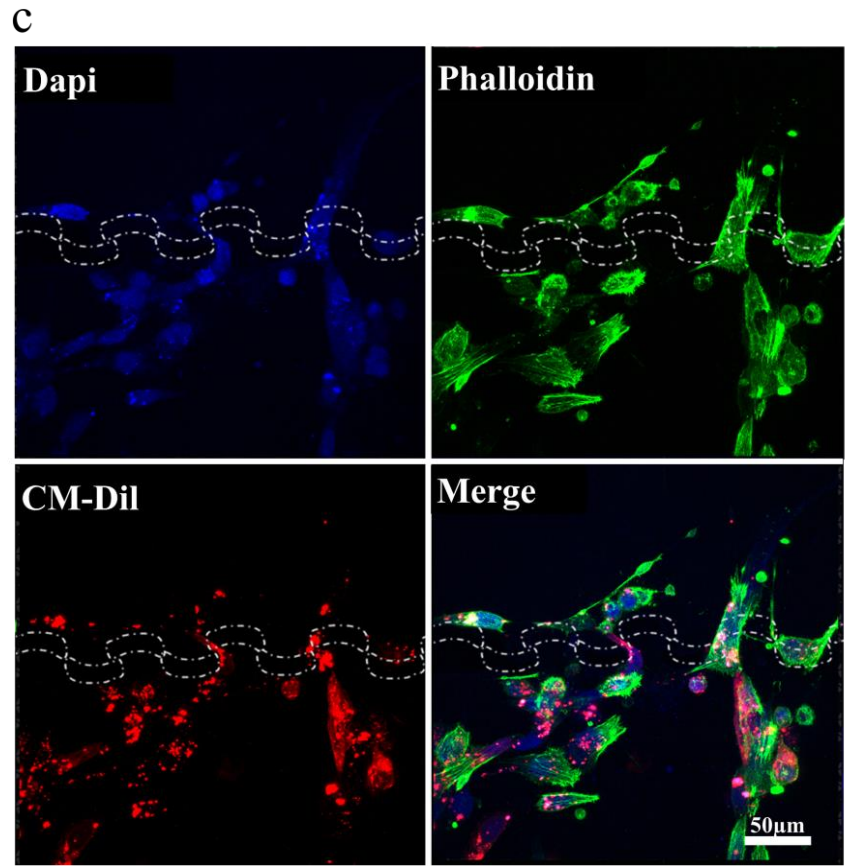

**Figure 6. Modeling the tissue invasion of cancer cells**. **a.** A schematic (top left) of the model and fluorescent microscopy images of the two compartments taken at 3, 6, 21, 24 and 30h. The images depict cancer cells (cell membrane labeled in red with CM-DiI) embedded in a matrigel matrix moving from one to the other compartment. **b.** Quantification of the number of migrated cancer cells up to 30h. Data are plotted as mean + SD, n=8. **c.** Confocal fluorescent images of the different channels were obtained using *Split-Channels* and recombining the images using *Merge-Channels*. Nuclei are stained in blue with DAPI, F-Actin is stained in green with phalloidin.

3636